\documentclass[a4paper,showpacs,prl,twocolumn]{revtex4}
\usepackage{amsfonts}
\usepackage{graphicx}
\usepackage{color}
\usepackage{amsmath}
\usepackage{amssymb}
\usepackage{latexsym}

\begin{document}

\title{Hysteresis and spin phase transitions in quantum wires in the integer
quantum Hall regime}
\author{S. Ihnatsenka and I. V. Zozoulenko}
\affiliation{Solid State Electronics, Department of Science and Technology (ITN), Link%
\"{o}ping University, 60174 Norrk\"{o}ping, Sweden}
\date{\today }

\begin{abstract}
We demonstrate that a split-gate quantum wire in the integer
quantum Hall regime can exhibit electronic transport hysteresis
for up- and down-sweeps of a magnetic field. This behavior is
shown to be due to phase spin transitions between two different
ground states with and without spatial spin polarization in the
vicinity of the wire boundary. The observed effect has a many-body
origin arising from an interplay between a confining potential,
Coulomb interactions and the exchange interaction. We also
demonstrate and explain why the hysteretic behavior is absent for
steep and smooth confining potentials and is present only for a
limited range of intermediate confinement slopes.
\end{abstract}

\pacs{73.21.Hb, 73.63.Nm, 73.43.-f, 73.43.Nq, 71.70.Di}

\maketitle

\textit{Introduction. } The effect of hysteresis represents fundamental
phenomena occurring in a variety of systems ranging from conventional
metallic ferromagnets \cite{Bertoni} to quantum resonant devices \cite%
{Goldman}. The hysteresis effect is usually associated with bistable
behavior where a system can undergo phase transitions or can have different
ground states. Recently, a new phenomenon of electronic transport hysteresis
has been observed in two-dimensional electron gas systems (2DEG) including
double-layered quantum well structures and conventional (one-layered)
modulation-doped heterostructures both in the regimes of the integer and the
fractional quantum Hall effects (IQHE and FQHE) \cite%
{Piazza,Cho,Eom,Smet,Zhu,Kronmuller,Poortere,Tutuc,Pan,Andy2}. The origin of
the hysteretic behavior in the double-layered structure was attributed to
the phase transition between two oppositely polarized ground states
localized in different layers \cite{Piazza}. The observations of hysteretic
behavior in the modulation-doped heterostructures in the FQHE suggest the
presence of a novel two-dimensional ferromagnetism where the spin-polarized
ground state competes with spin-unpolarized one \cite{Cho,Eom,Smet}, even
though the detailed many-body origin of the observed effect still remains
unclear. In contrast, the hysteresis reported in \cite{Zhu} in the IQHE
regime does not seem to require a many-body explanation and can be
understood in terms of a co-existence and a dynamical exchange of electrons
due to the presence a parallel conducting channel.

Theoretical investigations of hysteretic phenomena in 2DEG systems have
received less attention \cite{Rijkels,Manolescu1,Manolescu2,Chang}, being
mostly limited to periodically-modulated systems where hysteresis was
observed typically by varying the strength of the modulation (as opposed to
up- and down-sweeps of the magnetic fields used in the experiment).

In this Letter we predict transport hysteretic behavior for a quantum wire
geometry in the IQHE regime for up- and down-sweeps of the magnetic field.
We demonstrate that this behavior is related to the transition between two
phases corresponding respectively to two different ground states with
spatially spin-polarized and spatially spin-unpolarized edge channels in the
vicinity of the wire boundary. We also demonstrate and explain why the
hysteretic behavior is absent for steep and smooth confining potentials and
is present only for a limited range of intermediate confinement slopes. The
IQHE regime is usually associated with a one-electron description. We stress
that the observed effect has a many-body nature arising as an interplay
between a confining potential, the Coulomb interaction and the exchange
interaction.

We also note that the predicted hysteretic behavior not only sheds new light
on the structure of the edge states and spin transitions in the IQHE regime.
The predicted effect can occur in the leads of lateral nanostructures, which
might have important consequences for the magnetotransport of all lateral
quantum devices, particularly, for those exploiting spin polarized injection
and detection by means of the spatial separation of spins \cite{Andy,Andy2}.

\begin{figure*}[tb]
\includegraphics[scale=1.0]{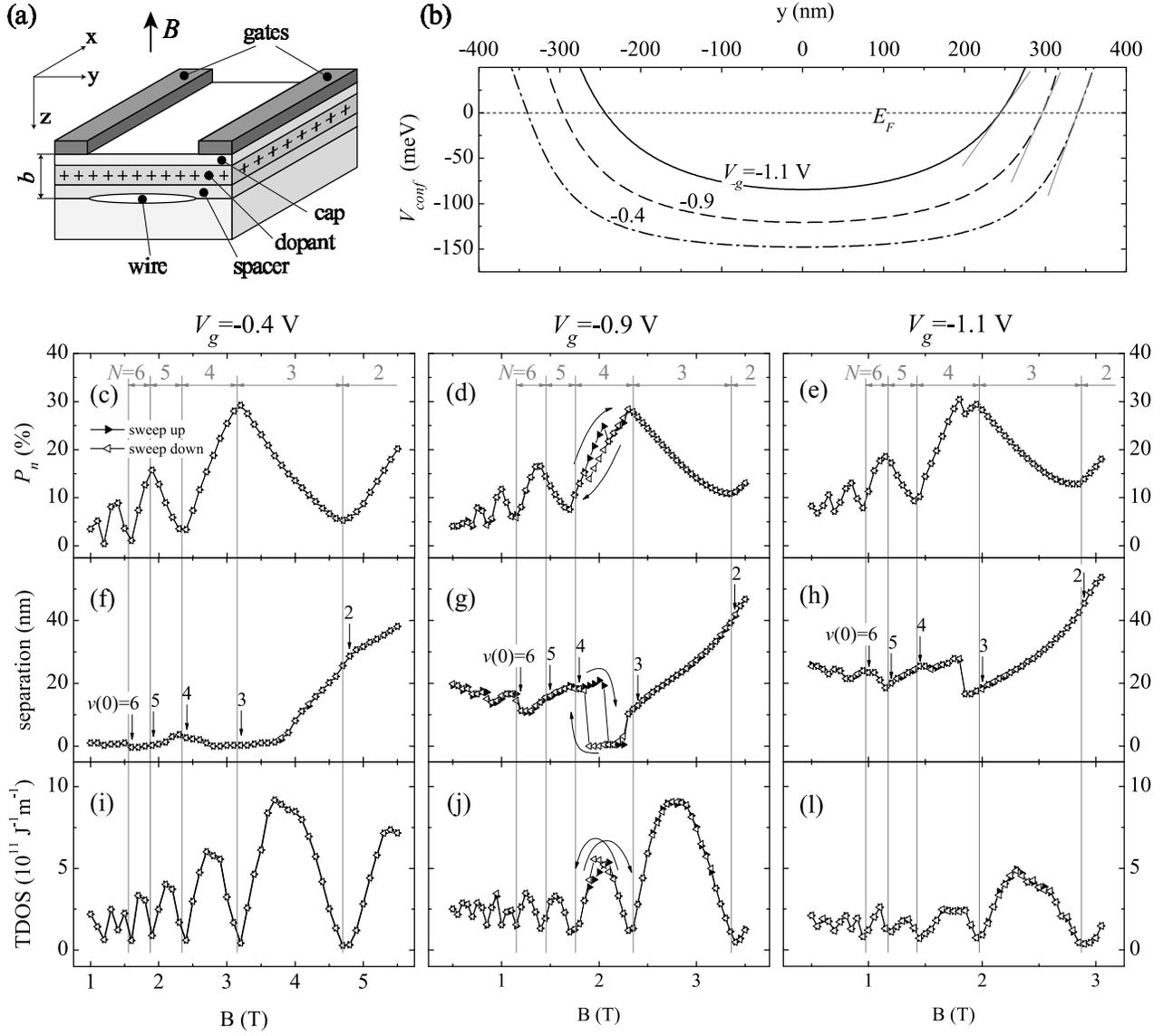}
\caption{(a) A schematic layout of a split-gate quantum wire in
perpendicular magnetic field. Parameters of the wire: the distance between
gates $w=700$ nm, the thickness of donor layer $d=26$ nm, the cap layer $%
c=14 $ nm, the distance between gates and the two-dimensional electron gas
(wire) $b_{2DEG}=50$ nm; the dopant concentration $1\cdot 10^{24}$ m$^{-3}$.
(b) Electrostatic potentials for the quantum wire for three gate voltages: $%
V_{g}=-0.4$ V, -0 (solid line), $V_{g}=-0.9$ V (dashed line), $V_{g}=-1.1$ V
(dash-dotted line). The tangents to the potential curves at $E_{F}$ indicate
the steepness of the potential near the boundary. (c)-(e) The charge density
spin polarization, (f)-(h) the spatial separation between the states $N=1$
and 2, and (i)-(l) the thermodynamical density of states for three gate
voltages from Fig. \protect\ref{fig:1} (b). $N$ and $\protect\nu (0)$
indicates the number of the occupied subbands and the local filling factor
in the middle of the wire. Temperature $T=1$ K.}
\label{fig:1}
\end{figure*}

In order to incorporate electron interaction and spin effects we use the
density functional theory (DFT) in the local spin density approximation \cite%
{ParrYang,TC}. The validity of this approximation for the 2DEG systems is
supported by the excellent agreement with the exact diagonalization and
variational Monte-Carlo calculations performed for few-electron systems \cite%
{Stephanie}. An important feature of our approach is that we start with a
lithographical geometry of the device (see Fig. \ref{fig:1} (a)) and do not
use any phenomenological parameters (such as charging constants, coupling
strengths etc.). In this respect, it is important to stress that the results
of the DFT-based transport modelling indicate that utilization of the
simplified approaches using phenomenological parameters and/or model
Hamiltonians might not always be reliable for theoretical predictions as
well as interpretations of experiments \cite{Martin}.

\textit{Model. }We consider a quantum wire in a perpendicular magnetic field
described by the Hamiltonian $H=\sum_{\sigma }H^{\sigma }$, $H^{\sigma
}=H_{0}+V^{\sigma }$, where $H_{0}$ is the kinetic energy in the Landau
gauge, $H_{0}=-\frac{\hbar ^{2}}{2m^{\ast }}\left\{ \left( \frac{\partial }{%
\partial x}-\frac{eiBy}{\hbar }\right) ^{2}+\frac{\partial ^{2}}{\partial
y^{2}}\right\} $, and $m^{\ast }=0.067m_{e}$ is the GaAs effective mass, see
Fig. \ref{fig:1} (a). The total confinement potential $V^{\sigma }$ for the
spin-up ($\sigma =+\frac{1}{2}$, $\uparrow $) and spin-down ($\sigma =-\frac{%
1}{2}$, $\downarrow $) electrons reads
\begin{equation}
V^{\sigma }=V_{conf}(y)+V_{H}(y)+V_{xc}^{\sigma }(y)+g\mu _{b}B\sigma ,
\label{total_potential}
\end{equation}%
where $V_{conf}(y)$ is the bare electrostatic confinement including
contributions from the the Schottky barrier $V_{Schottky}=0.8$ eV, the
split-gates \cite{Davies_gate} and the dopant layer \cite{Martorell} (for
the explicit expressions see \cite{Ihnatsenka}). The Hartree potential due
to the electron density $n(y)=\sum_{\sigma }n^{\sigma }(y)$ (including the
mirror charges) reads \cite{Ihnatsenka}
\begin{equation}
V_{H}(y)=-\frac{e^{2}}{4\pi \varepsilon _{0}\varepsilon _{r}}\int dy^{\prime
}n(y^{\prime })\ln \frac{\left( y-y^{\prime }\right) ^{2}}{\left(
y-y^{\prime }\right) ^{2}+4b^{2}},  \label{Hartree}
\end{equation}%
where $\varepsilon _{r}$ is the relative permittivity for GaAs $\varepsilon
_{r}=12.9$ and $b$ is the distance to the surface. $V_{xc}^{\sigma }(y)$ is
the exchange-correlation potential included within the framework of the
Kohn-Sham density functional theory \cite{ParrYang} in the local
spin-density approximation using the parameterization of Tanatar and
Cerperly \cite{TC} (see Ref. \onlinecite{Ihnatsenka} for explicit
expressions for $V_{xc}(y)$). The last term in Eq. (\ref{total_potential})
accounts for the Zeeman splitting with $\mu _{b}=\frac{e\hbar }{2m_{e}}$
being the Bohr magneton and the bulk $g$ factor of GaAs, $g=-0.44$. The
spin-resolved electron density is given by the retarded Green's function, $%
G^{\sigma }(y,y,E)$, $n^{\sigma }(y)=-\frac{1}{\pi }\Im \int dE\,G^{\sigma
}(y,y,E)f(E-E_{F}),$where $f(E-E_{F})$ is the Fermi-Dirac distribution
function. The Green's function of the wire as well as the electron and the
current densities are calculated self-consistently usin the technique
described in detail in Ref. \onlinecite{Ihnatsenka}. We calculate also the
thermodynamical density of states ($TDOS$) that reflects the structure of
magnetosubbands near the Fermi energy \cite{Davies_book},
\begin{equation}
TDOS=\sum_{\sigma }\int dE\,\rho ^{\sigma }(E)\left( -\frac{\partial
f(E-E_{F})}{\partial E}\right) ,  \label{tdos}
\end{equation}%
where the density of states $\rho ^{\sigma }(E)=-\frac{1}{\pi }\Im \int
dy\,G^{\sigma }(y,y,E)$ \cite{Datta_book}.

\begin{figure}[tb]
\includegraphics[scale=1.0]{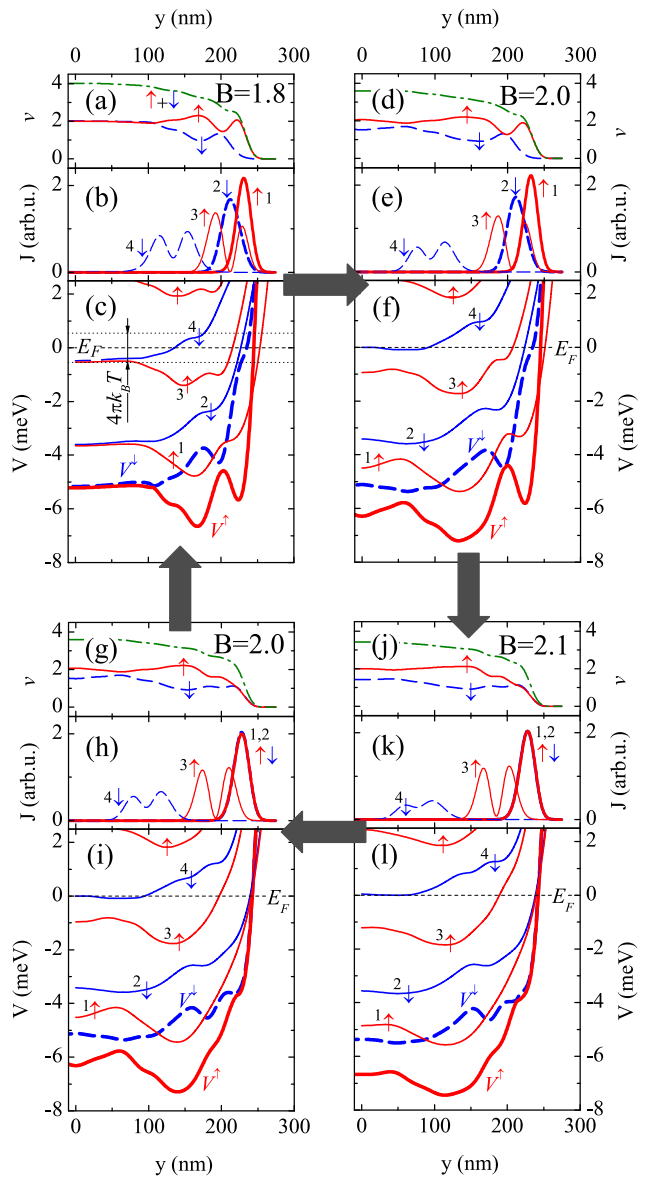}
\caption{The electron density profile, $\protect\nu =\frac{n(y)}{n_{B}}$,
current densities and magnetosubband structure for $V_{g}=-0.9$ V and four
values of magnetic field: (a)-(c) - $B=1.8$ T , (d)-(f) - $B=2.0$ T (sweep
up), (g)-(i) - $B=1.9$ T (sweep down), (j)-(k) - $B=2.1$ T. Big arrows show
the sweeping direction. Fat solid lines indicate the total confining
potentials $V^\protect\protect\sigma$; the integers 1-4 indicate the subband
number. Temperature $T=1$ K.}
\label{fig:2}
\end{figure}

\textit{Results and discussion. }Figure \ref{fig:1} shows the spin
polarization of the electron density,
$P_{n}=\frac{n_{1D}^{\uparrow
}-n_{1D}^{\downarrow }}{n_{1D}^{\uparrow }+n_{1D}^{\downarrow }}$ ($%
n_{1D}^{\sigma }=\int dy\,n^{\sigma }(y)$ is the one-dimensional density),
the spatial separation between the innermost (closest to the wire boundary)
edge states corresponding to the spin-up and spin-down subbands $N=1,$ 2 and
the $TDOS$ for three representative gate voltages. The density spin
polarization $P_{n}$ shows a distinct $1/B$-periodic looplike pattern, whose
periodicity is related to the subband depopulation (for a detailed analysis
of various aspects of the spin polarization and magnetosubband- and edge
state structure in quantum wires see Refs. \cite%
{Ihnatsenka,Ihnatsenka2,Ihnatsenka3}). In the present Letter we
concentrate at a pronounced hysteretic behavior for forward- and
reverse magnetic field
sweeps as shown in Fig. \ref{fig:1} (d),(g),(j) for the gate voltage $%
V_{g}=-0.9$ V. Note that the hysteretic behavior is present only
for some intermediate range of the gate voltage and is absent for
lower and larger voltages, see Fig. \ref{fig:1} (c),(f),(i) and
(e),(h),(l).

The key to understanding of the observed hysteresis effect is the spatial
polarization of the innermost edge states shown in Fig. \ref{fig:1} (g). For
the magnetic field $B\lesssim 1.8$ T (corresponding to the filling factor $%
\nu \gtrsim 4$), the spatial separation $d$ between $N=1$
(spin-up) and $N=2$ (spin-down) states is almost independent of
$B$ and is nearly constant. This spatial polarization is due to
the lifting of the spin degeneracy by the exchange interaction for
the case of a sufficiently smooth potential confinement
\cite{Dempsey}. For the magnetic field $B\gtrsim 2.1$ T ( $\nu
\lesssim 3$) the spatial separation $d$ gradually increases. This
behavior of $d$ can be traced back to the evolution of the
compressible strips in the Chklovsii \textit{et al. }
\cite{Chklovskii} model of the spinless electrons, where the width
of these strips, $w_{comp},$ monotonically increases with increase
of $B$. The exchange interaction is shown to affects dramatically
the compressible strips by suppressing them and inducing the
spatial separation of the spin states $d\approx w_{comp}$
\cite{Ihnatsenka2}.

In the field interval $1.8$ T$\lesssim B\lesssim 2.1$ T, the quantum wire
exhibits a bistable behavior with two distinct ground states with and
without the spatial spin polarization. The origin of the ground state with
the suppressed spatial spin polarization can be understood from an evolution
of the band structure of the quantum wire. For further analysis it is
important to emphasize that the Coulomb energy is dominant for the system at
hand such that the total electron density distribution $n(y)$ is practically
unaffected by the applied magnetic field \cite{Chklovskii}. Figures \ref%
{fig:2} (a)-(c) show the band structure and the current and electron
densities for $B=1.8$ T. The innermost spin-polarized states (denoted by 1
and 2 in (b)) are spatially separated due to the exchange interaction. The
spin-up state 3 belonging to the outermost subband partially overlaps with
the states 1 and 2. As the magnetic field increases the 3rd subband is
pushed up in the energy. As a result, the corresponding electron density is
redistributed towards the center of the wire, see Fig. \ref{fig:2} (e) for $%
B=2$ T, and the overlap between the states 2 and 3 increases. As
the state 3 moves away from the boundary, the density of the
remaining electrons has to be adjusted to keep the total density
$n(y)$ unchanged. This can be done only if the spin-down electrons
associated with the 2nd subband are redistributed towards the edge
of the wire. (Note that the Coulomb interaction is much stronger
than the exchange interaction separating states 1 and 2). This
redistribution leads to the collapse of the spatial spin
separation, see Fig. \ref{fig:2} (k) for $B=2.1$ T. This phase of
the system with the ground state without the spatial spin
polarization is preserved up to $\nu \approx 3$ for the sweep of
the magnetic field in the forward direction. For a higher field
(corresponding to the formation of compressible strips in the
Chklovskii \textit{et al.} model of spinless electrons) the
exchange interaction restores the spatial spin separation of the
order $d\approx w_{comp}$ \cite{Ihnatsenka2} as discussed above.

For the field sweep in the reverse direction, the spatial spin polarization
is restored for the magnetic field significantly lower ($\sim $1.8 T) than
that when the transition from the spin-polarized to the spin-unpolarized
phase takes place for the forward sweep ($\sim $2.1 T). The system shows a
memorization of the spin polarization, similar to a memorization of the
magnetization direction in ferromagnetic domains. Thus, in the above
interval ($1.8$ T$\lesssim B\lesssim 2.1$ T), the quantum wire exhibits a
bistable behavior where the system, depending on the history, can be in one
of the ground states (with or without the spatial spin polarization). The
bistable behavior of the quantum wire is manifest itself in the spin
polarization of the electron density $P_{n}$ (Fig. \ref{fig:1} (d)) and the
\textit{TDOS }(Fig. \ref{fig:1} (j)). The latter can be accessible via
magneto-capacitance \cite{Weiss} or magnetoresistance \cite{Berggren-dos}
measurements. The spin polarization of the electron density $P_{n}$ can be
probed directly using e.g. polarized photoluminescence spectra as recently
reported by Nomura and Aoyagi \cite{Nomura}. The predicted hysteretic
behavior can also be probed in transport measurements on lateral quantum
dots in the edge state regime involving spin polarized injection and
detection by means of the spatial separation of spins \cite{Andy,Andy2}.

Let us now discuss a condition for the hysteretic behavior to
occur. We demonstrate below that the key factor affecting the
hysteresis is a steepness of the electrostatic confinement near
the wire boundaries. [Note that application of more negative gate
voltage results in a potential
confinement with a less steep slope near the wire boundaries, see Fig. \ref%
{fig:1} (b)]. Figure \ref{fig:1} (f)] shows the spatial separation
$d$ for the applied voltage $V=-0.4$ V. In the magnetic field
region $3\lesssim \nu \lesssim 4$ (when the hysteresis is present
for $V=-0.9$ V), the spatial spin separation is practically
absent. This is because of a steep slope of the potential profile
such that the strength of the exchange interaction is not enough
to pull the spin species apart \cite{Dempsey}. As a result,
\textit{a phase corresponding to the spatially polarized ground
state in not present in wires with the steep walls and the
bistable behavior is not possible}. Opposite situation occurs for
the case of a smooth confinement, $V=-1.1$ V, see Fig. \ref{fig:1}
(h), when the ground state can be only spatially polarized in the
above field interval ($3\lesssim \nu \lesssim 4$) . An inspection
of the corresponding band structure calculated in the Hartree
approximation for the spinless electrons (which well corresponds
to the Chklovskii \textit{et al. }model \cite{Ihnatsenka2,Ando})
reveals an onset of a formation of the compressible strips. [The
compressible strips are more easily formed in a wire with more
smooth confinement \cite{Chklovskii}]. As discussed above, the
compressible strips are suppressed by the exchange interaction
\cite{Ihnatsenka2},
which leads to the spatial separation of the spin states $d\approx w_{comp}$%
. \textit{This results in the absence of the spin-unpolarized
phase (and thus the hysteretic behavior) in wires with a smooth
confinement}.

It should be stressed that the results and conclusions reported above are
neither specific nor limited to the particular parameters of the quantum
wire of Fig. \ref{fig:1}. Similar hysteretic behavior has been detected in
different quantum wires. Note that in some cases a less pronounced
hysteresis (of the same origin as above due to the collapse of the spatial
spin separation) has been also seen in a vicinity of $\nu \approx 6$. The
parameters of the quantum wire considered in this Letter have been chosen so
as to motivate a possible experiment in which a gate voltage sweep on the
same device can span all characteristic regions of the steep, the
intermediate and the smooth confinement.

In conclusion, we have demonstrated that a quantum wire in the IQHE regime
exhibits a hysteretic behavior and spin phase transitions arising as an
interplay between a confining potential, the Coulomb interactions and the
exchange interaction.

S. I. acknowledges a financial support from the Swedish Institute.


\begin{thebibliography}{99}
\bibitem{Bertoni} G. Bertoni, \textit{Hysteresis and magnetism} (Academic
press, New Yoork, 1998).

\bibitem{Goldman} V. J. Goldman, D. C. Tsui, and J. E. Cunningham, Phys.
Rev. Lett. \textbf{58}, 1256 (1987).


\bibitem{Piazza} V. Piazza, V. Pellegrini, F. Beltram, W. Wegscheider, T.
Jungwirth, A. H. MacDonald, Nature \textbf{402}, 638 (1999).

\bibitem{Cho} H. Cho, J. B. Young, W. Kang, K. L. Campman, A. C. Gossard, M.
Bichler, and W. Wegscheider, Phys. Rev. Lett. \textbf{81}, 2522 (1998)

\bibitem{Eom} J. Eom, H. Cho, W. Kang, K. L. Campman, A. C. Gossard, M.
Bichler, and W. Wegscheider, Science \textbf{289}, 2320 (2000).

\bibitem{Smet} J. H. Smet \textit{et al.}, Phys. Rev. Lett.\textbf{\ 86},
2412 (2001).

\bibitem{Zhu} J. Zhu \textit{et al.}, Phys. Rev. B \textbf{61}, R13 361
(2000).

\bibitem{Kronmuller} S. Kronmuller \textit{et al.}, Phys. Rev. Lett. \textbf{%
81}, 2526 (1998).

\bibitem{Poortere} E. P. De Poortere, E. Tutuc, S. J. Papadakis, M.
Shayegan, Science \textbf{290}, 1546 (2000).

\bibitem{Tutuc} E. Tutuc \textit{et al.}, Phys. Rev. B \textbf{68},
201308(R) (2003)

\bibitem{Pan} W. Pan, J. L. Reno and J. A. Simmons, Phys. Rev. B \textbf{71}, 153307 (2005).

\bibitem{Andy2} M. Pioro-Ladriere \textit{et al.}, Phys. Rev. B \textbf{73}, 075309 (2006).


\bibitem{Rijkels} L. Rijkels and G. E. W. Bauer, Phys. Rev. B \textbf{50},
8629 (1994).

\bibitem{Manolescu1} A. Manolescu and V. Gudmundsson, Phys. Rev. B \textbf{59%
}, 5426 (1999).

\bibitem{Manolescu2} A. Manolescu and V. Gudmundsson, Phys. Rev. B \textbf{61%
}, R7858 (2000).

\bibitem{Chang} M.-C. Chang and M.-F. Yang, Phys. Rev. B \textbf{64}, 073302
(2001).


\bibitem{Andy} M. Ciorga \textit{et al.},
Phys. Rev. B. 61, R16315 (2000); A. S. Sachrajda, \textit{et al.},
Physica E 10\}, 493 (2001).


\bibitem{ParrYang} R. G. Parr and W. Yang, \textit{Density-Functional Theory
of Atoms and Molecules}, (Oxford Science Publications, Oxford, 1989).

\bibitem{TC} B. Tanatar and D. M. Ceperley, Phys. Rev. B \textbf{39}, 5005,
(1989).

\bibitem{Stephanie} S. M. Reimann and M. Manninen, Rev. Mod. Phys. \textbf{74%
}, 1283 (2002); E. R\"{a}s\"{a}nen \textit{et al.},
Phys. Rev B \textbf{67}, 235307 (2003). M. Borgh \textit{et al.}, Intl. J.
Quant. Chem. \textbf{105}, 817 (2005).

\bibitem{Martin} M. Evaldsson and I. V. Zozoulenko, Phys. Rev. B \textbf{73}%
, 035319 (2006).


\bibitem{Davies_gate} J. H. Davies, I. A. Larkin, and E. V. Sukhorukov, J.
Appl. Phys. \textbf{77}, 4504 (1995).

\bibitem{Martorell} J. Martorell, H. Wu and D. W. L. Sprung, Phys. Rev. B
\textbf{50}, 17298 (1994).

\bibitem{Ihnatsenka} S. Ihnatsenka and I. V. Zozoulenko, Phys. Rev. B
\textbf{73}, 075331 (2006).

\bibitem{Davies_book} J. Davies, \textit{The Physics of Low-Dimensional
Semiconductors}, (Cambridge University Press, Cambridge, 1998).

\bibitem{Datta_book} S. Datta, \textit{Electronic Transport in Mesoscopic
Systems}, (Cambridge University Press, Cambridge, 1997).

\bibitem{Ihnatsenka2} S. Ihnatsenka and I. V. Zozoulenko, Phys. Rev. B
\textbf{73}, 155314 (2006).

\bibitem{Ihnatsenka3} S. Ihnatsenka and I. V. Zozoulenko, Phys. Rev. B, to
be published (cond-mat/0605008).

\bibitem{Dempsey} J. Dempsey, B. Y. Gelfand, and B. I. Halperin, Phys. Rev.
Lett. \textbf{70}, 3639 (1993).

\bibitem{Chklovskii} D. B. Chklovskii, B. I. Shklovskii, and L. I. Glazman,
Phys. Rev. B \textbf{46}, 4026 (1992); D. B. Chklovskii, K. A. Matveev, and
B. I. Shklovskii, Phys. Rev. B \textbf{47}, 12605 (1993).

\bibitem{Weiss} D. Weiss, C. Zhang, R. R. Gerhardts, K. v. Klitzing, G.
Weimann, Phys. Rev. B \textbf{39}, 13020 (1989).

\bibitem{Berggren-dos} K.-F. Berggren, G. Roos, and H. van Houten, Phys.
Rev. B \textbf{37}, 10118 (1998).

\bibitem{Nomura} S. Nomura and Y. Aoyagi, Phys. Rev. Lett. \textbf{93},
096803 (2004).

\bibitem{Ando} T. Suzuki and T. Ando, Physica B \textbf{249-251}, 415 (1998).
\end{thebibliography}
\end{document}